# X-ray Properties of Young Early Type Galaxies: I. X-ray Luminosity Function of LMXBs

Dong-Woo Kim and Giuseppina Fabbiano

Smithsonian Astrophysical Observatory, 60 Garden Street, Cambridge, MA 02138

(Thursday, August 05, 2010)

**ABSTRACT**

We have compared the combined X-ray luminosity function (XLF) of LMXBs detected in Chandra observations of young, post-merger elliptical galaxies, with that of typical old elliptical galaxies. We find that the XLF of the 'young' sample does not present the prominent high luminosity break at $L_x > 5 \times 10^{38}$ erg s$^{-1}$ found in the old elliptical galaxy XLF. The 'young' and 'old' XLFs differ with a $3\sigma$ statistical significance (with a probability less than 0.2% that they derive from the same underlying parent distribution). Young elliptical galaxies host a larger fraction of luminous LMXBs ($L_x > 5 \times 10^{38}$ erg s$^{-1}$) than old elliptical galaxies and the XLF of the young galaxy sample is intermediate between that of typical old elliptical galaxies and that of star forming galaxies. This observational evidence may be related to the last major/minor mergers and the associated star formation.

*Subject headings*: galaxies: elliptical and lenticular – X-rays: binaries –X-rays: galaxies

## 1. Introduction

Merging has been suggested to be a key factor in the formation of elliptical galaxies (e.g., Toomre & Toomre 1972). However, the general lack of observational evidence for post-merger intermediate-age systems has made it difficult to test the merger hypothesis (e.g., King 1977). The stellar age of an elliptical galaxy can be measured with three different methods: a spectroscopic age estimator from comparison of observed spectra with spectral/population synthesis models (e.g., Trager et al. 2000; Terlevich and Forbes 2002); a dynamical age estimator from morphological fine structures (e.g., Schweizer and Seitzer 1992); and a photometric age estimator from the colors of globular clusters (e.g., Whitmore et al. 1997; Larsen et al. 2001). These methods, although not always giving consistent results, have provided a handful of elliptical galaxies with relatively young stellar ages (< 5 Gyr). As stars are fading out after the merger-induced star formation, obvious signatures of star formation such as OB stars and SNe disappear well before a few 100 Myr. It is the 1-5 Gyr old post-merger system which carries the key observational evidence of the missing link between Antennae-like merging systems and typical old elliptical galaxies.

Although these optical age measurements dramatically increase our understanding of the formation and evolution of elliptical galaxies, there are still a number of issues in the optical data analysis and interpretation, such as correlated errors, aperture bias and incomplete model calibrations (e.g., Proctor et al. 2004; Thomas et al. 2005). Furthermore the optically measured age is often luminosity-weighted rather than mass-weighted under the assumption of a single stellar population. More realistic analysis may result in somewhat different ages (e.g., Idiart et al. 2007; Rogers et al. 2009; Trager and Somerville 2009).

In this paper, we report, for the first time, an X-ray signature of age. We have investigated the Chandra data by carefully selecting a sample of young elliptical galaxies and comparing the X-ray luminosity function (XLF) of the low-mass X-ray binaries (LMXB) in these galaxies with that of typical old ellipticals, and we report the results here. In an accompanying paper (Kim et al. 2010 in preparation), we will address distinct metal abundance ratios ($Fe$ to $\alpha$-elements) in the hot ISM of young and old elliptical galaxies.

This paper is organized as follows. In Section 2, we describe our sample selection, Chandra observations and data reduction techniques. In Section 3, we quantitatively compare the completeness-corrected XLFs of young and old elliptical galaxy samples. We also perform a straightforward statistical test to show different fractions of luminous LMXBs between the two samples. In Section 4, we discuss the implications of our results for the age effect on LMXB XLF and summarize our conclusions.

## 2. Sample Selection and Chandra Observations

We selected young elliptical galaxies which were observed with Chandra, according with the following criteria: (1) young age (< 5 Gyr, taken from Trager et al. 2000, Terlevich & Forbes 2002 and Thomas et al. 2005); (2) proximity (≤25 Mpc, taken from Tonry et al. 2001); (3) luminosity ($M_B$ ≤ -20.0 mag). We impose the 2nd and 3rd conditions to ensure that a large number of LMXBs can be detected. We further impose an additional condition of (4) the LMXB detection limit to be $L_X < 10^{38}$ erg s$^{-1}$, so that the luminosity range of the known XLF break at $L_X \cong 5 \times 10^{38}$ erg s$^{-1}$ (Kim & Fabbiano 2004; Gilfanov 2004) can be accurately observed in both young and old samples. NGC 3585 and NGC 3923 were targeted in the AO9 Chandra observing cycle by our team specifically for this purpose and the remaining Chandra data were obtained from the Chandra archive (see Table 2).

For comparison, we have selected a sample of old elliptical galaxies by applying similar criteria, except (1) being old (> 7 Gyr). Additionally we exclude gas-rich old elliptical galaxies which are mostly cD type cluster/group dominant elliptical galaxies where faint LMXBs could easily be confused with hot gas substructures. We note that there is no gas-rich galaxy among the young elliptical galaxy sample (see Kim et al. 2010 in preparation; this lack of gaseous halos in young ellipticals was first noted by Fabbiano & Schweizer 1995).

Table 1a Young Elliptical Galaxies

| name  | T    | d (Mpc) | B (mag) | $M_B$ (mag) | K (mag) | $L_K$ ($10^{10}$ erg s$^{-1}$) | $D_{25}$ (arcmin) | age1 | age2 (Gyr) | age3 | $S_N$ |
|-------|------|---------|---------|-------------|---------|-------------------------------|-------------------|------|------|------|-------|
| N0720 | -5.0 | 27.6    | 11.13   | -21.08      | 7.27    | 20.31                         | 4.6 x 2.3         | 3.4  | 4.5  | 5.4  | 2.2   |
| N1316 | -2.0 | 21.4    | 9.40    | -22.26      | 5.58    | 57.71                         | 12.0 x 8.5        | 3.4  |      | 3.2  | -     |
| N3377 | -5.0 | 11.2    | 11.07   | -19.18      | 7.44    | 2.85                          | 5.2 x 3.0         | 4.1  | 3.7  | 3.6  | 2.4   |
| N3585 | -5.0 | 20.0    | 10.64   | -20.86      | 6.70    | 17.97                         | 4.6 x 2.5         | 3.1  |      |      | -     |
| N3923 | -5.0 | 22.9    | 10.62   | -21.18      | 6.50    | 28.32                         | 5.8 x 3.8         |      |      | 3.3  | 6.8   |
| N4125 | -5.0 | 23.9    | 10.67   | -21.22      | 6.86    | 21.88                         | 5.8 x 3.2         | 5.0  |      |      | -     |
| N4382 | -1.0 | 18.4    | 9.99    | -21.34      | 6.14    | 25.47                         | 7.0 x 5.4         | 1.6  |      |      | 1.3   |

Table 1b Old Elliptical Galaxies

| name  | T    | d (Mpc) | B (mag) | $M_B$ (mag) | K (mag) | $L_K$ ($10^{10}$ erg s$^{-1}$) | $D_{25}$ (arcmin) | age1 | age2 (Gyr) | age3 | $S_N$ |
|-------|------|---------|---------|-------------|---------|-------------------------------|-------------------|------|------|------|-------|
| N1404 | -5.0 | 20.9    | 10.98   | -20.63      | 6.82    | 17.70                         | 3.3 x 2.9         | 5.9  | 9.0  |      | 3.2   |
| N3379 | -5.0 | 10.5    | 10.18   | -19.94      | 6.27    | 7.45                          | 5.3 x 4.7         | 9.3  | 8.6  | 10.0 | 1.2   |
| N4278 | -5.0 | 16.0    | 10.97   | -20.06      | 7.18    | 7.42                          | 4.0 x 3.8         | 10.7 | 0.0  | 12.0 | 6.9   |
| N4374 | -5.0 | 18.3    | 10.01   | -21.31      | 6.22    | 23.52                         | 6.4 x 5.6         | 11.8 | 12.2 | 12.8 | 5.2   |
| N4472 | -5.0 | 16.2    | 9.33    | -21.73      | 5.39    | 39.58                         | 10.2 x 8.3        | 8.5  | 7.9  | 9.6  | 5.4   |
| N4552 | -5.0 | 15.3    | 10.57   | -20.36      | 6.72    | 10.30                         | 5.1 x 4.6         | 9.6  | 10.5 | 12.4 | 2.8   |
| N4697 | -5.0 | 11.7    | 10.07   | -20.28      | 6.36    | 8.42                          | 7.2 x 4.6         | 8.2  | 8.9  | 8.3  | 2.5   |

Our samples of young and old elliptical galaxies are listed in Table 1. The morphological type *T*, *B*-band magnitude, and $D_{25}$ ellipse are taken from RC3, the *K*-band magnitude from the 2MASS through NED, and the distance from Tonry et al. (2001). We list three age estimates from Trager et al. (2000)

Terlevich & Forbes (2002) and Thomas et al. (2005). We add another well-known young elliptical galaxy, NGC 4125 (see Schweizer and Seitzer 1992 and Fabbiano and Schweizer 1995). We take the age of NGC 4125 from Schweizer and Seitzer (1992), based on the optical color indices and fine structures. In the last column we list the GC specific frequency ($S_N = N_{GC} \times 10^{0.4(M_V + 15)}$) taken from Peng et al. (2008) and Ashman & Zepf (1998). For galaxies listed in both, we take $S_N$ from Peng et al. (2008), because the HST result is more reliable in identifying globular clusters and in reducing contaminations than those based on the ground-based observations.

We note that all galaxies are classified as *E* (or *T*=-5) in RC3, except NGC 1316 and NGC 4382. NGC 1316 is a giant elliptical galaxy in the Fornax cluster (also a radio source known as Fornax A). Because of evidence of recent mergers (e.g., Schweizer 1980), it is often classified as a peculiar *E* or *S0*. NGC 4382 is *S0* peculiar, where the inner region resembles a normal *E5* (Sandage & Bedke 1994) and the outer region shows twisted isophotes (e.g., Fisher 1997).

```
               Table 2a Chandra Observations of Young Ellipticals
___________________________________________________________________________________

  Name    Obsids                      Observation dates     Exp (ksec) PIs              ref
___________________________________________________________________________________

 N0720   492, 7372, 7062, 8448, 8449  2000.10.12-2006.10.12   127.8    Garmire/Humphrey  a
 N1316   2022                         2001.04.17               24.1    Kim               b
 N3377   2934                         2003.01.06               39.2    Irwin
 N3585   2078, 9506                   2001.06.03-2008.03.11    90.2    Bregman/Kim       c
 N3923   1563, 9507                   2001.06.14-2008.04.11    93.4    Murray/Kim        c
 N4125   2071                         2001.09.09               60.7    White
 N4382   2016                         2001.05.29               39.0    Sarazin           d
___________________________________________________________________________________

                Table 2b Chandra Observations of Old Ellipticals
___________________________________________________________________________________

  Name    Obsids                      Observation date      Exp (ksec) PIs              ref
___________________________________________________________________________________

 N1404   2942                         2003.02.13               28.7    Irwin             e
 N3379   1587, 7073-7076              2001.02.13-2007.01.10   324.2    Murray/Fabbiano   f,g
 N4278   4741, 7077-7080              2005.02.03-2007.04.20   458.0    Irwin/Fabbiano    h
 N4374    803                         2000.05.19               27.1    Finoguenov        i
 N4472    321                         2000.06.12               32.1    Mushotzky         j,k,l
 N4552   2072                         2001.04.22               47.9    White             m,n
 N4697   4727-4730                    2003.12.26-2004.08.18   132.0    Sarazin           o,p
___________________________________________________________________________________
```

a. Jeltema, T. E., et al. 2002 ApJ, 585, 756
b. Kim, D.-W. & Fabbiano, G. 2003 ApJ, 586, 826
c. this paper
d. Sivakoff, G. R., et al. 2003, ApJ, 599, 218
e. Machacek, M. et al. 2005, ApJ, 621, 663
f. David, L. P., et al. 2005, ApJ, 635, 1053
g. Brassington, N. J. et al. 2008, ApJS, 172, 142
h. Brassington, N. J. et al. 2009, ApJS, 181, 605

All Chandra observations were taken with the *S3* (back-illuminated) chip of the Advanced CCD Imaging Spectrometer (ACIS, Garmire 1997). Some galaxies were observed multiple times, with individual exposures ranging from 10 to 100 ks. Observation dates and combined net exposure times (after removing periods of background flares) are summarized in Table 2. In all the observations used in this study, the ACIS temperature was -120 C. We did not use a small number of old observations taken before January 2000 with detector temperature of -110 C, because of the relatively large uncertainty in calibrating the detector characteristics (http://cxc.harvard.edu/cal/Acis/). Also listed in Table 2 are principal investigators of the original observations and previous works focused on individual galaxies. Some galaxies were also analyzed in previous works for different scientific goals, e.g., Irwin et al. (2003, 2004), Kim & Fabbiano (2004), Sivakoff et al. (2007) and Humphrey & Buote (2008).

The ACIS data were uniformly reduced in a similar manner as described in Kim & Fabbiano (2003) with a custom-made pipeline (XPIPE), specifically developed for the Chandra Multi-wavelength Project (ChaMP; Kim et al. 2004). Starting with the CXC pipeline level 2 products, we apply *acis_process_event* available in CIAO v3.4 with up-to-date calibration data, e.g., time-/position-dependent gain and QE variation. For targets with multiple observations, we re-project individual observations to a common tangent point and combine them by using *merge_all* available in the CIAO contributed package (http://cxc.harvard.edu/ciao/threads/combine/).

The X-ray point sources were detected using CIAO *wavdetect*. We set the significance threshold to be $10^{-6}$, which corresponds approximately to one false source per chip and the exposure threshold to be 10% using an exposure map. The latter was applied to reduce the false detections often found at the chip edge. The performance and limitations of *wavdetect* are well understood and calibrated by extensive simulations (e.g., Kim et al. 2004; Kim, M. et al. 2007).

To measure the X-ray flux and luminosity (in 0.3-8 keV), we take into account the temporal and spatial QE variation (http://cxc.harvard.edu/cal/Acis/Cal_prods/qeDeg/) by calculating the energy conversion factor (ECF = ratio of flux to count rate) for each source in each observation. We assume a power-law spectral model with a photon index of $\Gamma$ = 1.7 (e.g., Irwin et al. 2003) and Galactic $N_H$ (see Table 1). To calculate the X-ray flux of sources detected in the merged data, we apply an exposure-weighted mean ECF. This will generate a flux as if the entire observations were done in one exposure, but with a variable detector QE as in the real observations.

## 3. X-ray Luminosity Function of LMXBs

To build the XLFs, we used point sources detected within the $D_{25}$ ellipse (the size and position angle are given in Table 1). Although some X-ray sources outside the $D_{25}$ ellipse may be associated with the galaxy, we excluded them to reduce the contamination by interlopers. We also excluded sources located near the galactic centers (R < 10″), because of large photometric errors and difficult incompleteness correction. Applying these criteria, we use 358 and 395 sources detected in the young and old samples, respectively.

To test whether the LMXB XLFs are different between young and old samples, we built completeness-corrected XLFs of two samples and quantitatively compare them in section 3.1. Then, guided by these results we also directly compared the number of high/low luminosity LMXBs by applying a straightforward statistical test in section 3.2.

## 3.1 Luminosity Function

To determine the XLFs accurately, it is most critical to correct for incompleteness (see Kim & Fabbiano 2003, 2004 hereafter KF04). Without this correction, the XLF would appear flattened at the lower luminosities where the detection is not complete, causing an artificial break. Following KF04, we performed extensive simulations to generate incompleteness corrections: we simulated 20,000 point sources using **MARX** (http://space.mit.edu/ASC/MARX/), added them one by one to the observed image and then determined whether the added source is detected. In the simulations, we assumed a typical LMXB XLF differential slope of β=2 (KF04) where β is defined in the differential form. We note that the adopted XLF slope does not significantly affect the results, because the correction is determined by the ratio of the number of input sources to that of detected sources at a given $L_X$ (see also Kim and Fabbiano 2003). As shown in Brassington et al. 2008, 2009 (see also Kim E. et al. 2006), the radial distribution of LMXBs closely follows that of the optical halo light. Therefore, we adopted an $r^{-1/4}$ law for the radial distribution of the LMXBs. Even if the radial distribution of LMXBs deviated from that of the stellar distribution, the effect would be minimal, because we do not use LMXBs from the central regions (r<10″) where the uncertainty in the incompleteness correction obtained by using different radial profiles would be most significant.

To build the XLF of LMXBs in the two samples separately, we combined all LMXBs in each sample after correcting for the incompleteness in each galaxy. To determine the XLF shape parameters, we fitted the bias-corrected, combined XLF in a differential form with (a) a single power-law, and (b) a broken power-law. We applied both $\chi^2$ and Cash statistics, using **sherpa** available in the CIAO package. To properly apply the $\chi^2$ statistic we selected the $L_X$ bin size, $\delta\log(L_X)$ = 0.1, so that there is a minimum of 10 sources in each $L_X$ bin. At high luminosities (Lx > 1.2 x $10^{39}$ and Lx > 6.3 x $10^{38}$ erg s$^{-1}$ for young and old samples, respectively), we further co-added a few bins to satisfy this condition. We apply the Gehrels variance function for the error calculation (Gehrels 1986). The Cash statistic (or cstat) utilizes a maximum likelihood function and can be applied regardless of the number of sources in each bin. In this case, we use the same bin size, $\delta\log(L_X)$ = 0.1, but without further adding bins at the higher Lx. The **sherpa** cstat also provides an approximate goodness-of-fit measure (http://cxc.harvard.edu/sherpa/

statistics/#cstat), i.e., the observed statistic divided by the number of degrees of freedom which should be close to 1 for good fits. Because in the Cash statistic the counts are sampled from the Poisson distribution in each bin, we cannot fit the corrected XLF, because the bias corrected sources counts would no longer be Poissonian. Instead, we fitted the uncorrected XLF with the modified model, which is divided by the correction factor. When we plot the XLF, the correction factor is multiplied back to the model. Both statistics result in consistent parameters within the error.

We present fitting results in Table 3 and show the combined, bias-corrected XLFs for the young and old elliptical galaxy samples in Figure 1. The differential XLF is plotted in the form of $dN/d\ln L_X$ as a function of $L_X$ (instead of $dN/dL_X$ vs. $L_X$). In this form, the slope, if a single power-law is applied, will be the same as that of the cumulative XLF so that the XLF is easily visualized and compared (e.g., Voss & Gilfanov 2007; Kim et al. 2009). As clearly seen in Figure 1, the XLFs of the young and old elliptical galaxy samples differ: while the two XLFs look similar at lower X-ray luminosities ($L_X \lesssim 5 \times 10^{38}$ erg s$^{-1}$), the XLF of the young sample has a considerably flatter slope than that of the old sample at higher X-ray luminosities ($L_X \gtrsim 5 \times 10^{38}$ erg s$^{-1}$).

For the young elliptical galaxy sample, a single power-law well reproduces the observed XLF with reduced $\chi^2$ = 0.52 for 14 degrees of freedom (92% probability to reject a null hypothesis) and reduced C statistic of 1.1 for 19 degrees of freedom (34% probability). Instead, for the old galaxy sample, a single power-law does not fit the observed XLF which shows a clear deficit at high luminosities compared to the best-fit single power-law (see the bottom left panel in Figure 1). In this case, reduced $\chi^2$ = 1.36 for 12 degrees of freedom (17% probability) and reduced C statistic of 1.71 for 13 degrees of freedom (5% probability). We note that in both samples, the best fit slope is $\beta$ = 1.9-2 which is consistent with that of KF04 ($\beta$ = 2.1 ± 0.1) for the same model.

Table 3. XLF parameters

Young Elliptical Galaxies

|  | Cstat | $\chi^2$ stat |
|---|---|---|
| <single power-law fit> | | |
| β | 2.01  -0.05 +0.05 | 2.04  -0.06 +0.07 |
| statistic | 1.10 (21.0/19) | 0.52 (7.28/14) |
| Probability | 0.340 | 0.923 |
| <broken power-law fit> | | |
| β1 | 1.91  -0.05 +0.06 | 1.94  -0.07 +0.08 |
| β2 | 2.71  -0.29 +0.33 | 2.82  -0.44 +0.78 |
| Lx(break) | 7.77  -1.38 +2.43 | 8.75  -2.41 +4.49 |
| statistic | 0.91 (15.4/17) | 0.35 (4.24/12) |
| Probability | 0.564 | 0.980 |

Old Elliptical Galaxies

| | Cstat | $\chi^2$ stat |
|---|---|---|
| <single power-law fit> | | |
| β | 1.92 −0.05 +0.06 | 1.99 −0.06 +0.06 |
| statistic | 1.71 (22.3/13) | 1.36 (16.33/12) |
| Probability | 0.051 | 0.177 |
| <broken power-law fit> | | |
| β1 | 1.78 −0.08 +0.08 | 1.78 −0.08 +0.08 |
| β2 | 3.73 −0.68 +0.89 | 9.99 −6.63 --- |
| Lx(break) | 5.18 −0.82 +0.81 | 6.92 −2.35 +1.60 |
| statistic | 0.75 (8.2/11) | 0.39 (3.90/10) |
| Probability | 0.692 | 0.952 |

If we apply a broken power-law, the difference between two samples is clearly reflected in two parameters, the slope at high luminosities and the break luminosity. While the slope (β = 1.8-1.9) at low luminosities (below the break) is consistent in the two samples within the error, the slope at high luminosities (above the break) is flatter in the young sample than in the old sample. Also noticeable is that even if both XLFs break with steeper slopes at high luminosities, the break luminosity is higher in the young sample than in the old sample, again suggesting that the young elliptical galaxies host more luminous LMXBs. Using $\chi^2$ statistic results, we perform the F-test to quantitatively determine whether the broken power-law is required. The F-test significance is 0.25 and 0.028 for the young and old samples, respectively, indicating that the broken power-law significantly improve the XLF fit for the old sample, but may not for the young sample. In summary, we find a considerably higher fraction of luminous LMXBs in the young galaxy sample than in the old sample.

## 3.2 Luminosity Distribution

To illustrate the difference between young and old elliptical galaxies in a more straightforward manner, we directly compare the number of luminous LMXBs in the two samples. Dividing LMXBs into two groups at $L_X = 5 \times 10^{38}$ erg s$^{-1}$, which corresponds to the XLF break luminosity (KF04 and Gilfanov 2004), we measure the fraction ($F_{LL}$) of luminous LMXBs with $L_X > 5 \times 10^{38}$ erg s$^{-1}$ out of all LMXBs with $L_X > 10^{38}$ erg s$^{-1}$,

$F_{LL} = N_{LMXB} (L_X > 5 \times 10^{38}$ erg s$^{-1}) / N_{LMXB} (L_X > 10^{38}$ erg s$^{-1})$.

In Table 4 we list the number of LMXBs in different luminosity bins. As described above, our sample galaxies were observed with Chandra for an exposure time long enough to detect LMXBs with $Lx \gtrsim 10^{38}$ erg s$^{-1}$ (at 90%), so the incompleteness correction for $F_{LL}$ is only minor.

Table 4 Number of LMXBs in different Lx bins

| | $L_X^a$ = 1-5 | 5-10 | >10 | >20[b] | $F_{LL}$ |
|---|---|---|---|---|---|
| **Young Ellipticals** | | | | | |
| N0720 | 35 | 10 | 3 | 0 | 0.27 |
| N1316 | 52 | 10 | 6 | 1 | 0.24 |
| N3377 | 4 | 1 | 0 | 0 | 0.20 |
| N3585 | 18 | 2 | 0 | 0 | 0.10 |
| N3923 | 39 | 8 | 3 | 3 | 0.22 |
| N4125 | 17 | 3 | 1 | 0 | 0.25 |
| N4382 | 23 | 5 | 2 | 1 | 0.23 |
| | 188 | 39 | 15 | 5 | 0.22 |
| **Old Ellipticals** | | | | | |
| N1404 | 13 | 2 | 1 | 0 | 0.19 |
| N3379 | 4 | 1 | 0 | 0 | 0.20 |
| N4278 | 20 | 3 | 0 | 0 | 0.13 |
| N4374 | 38 | 3 | 2 | 1 | 0.12 |
| N4472 | 78 | 6 | 1 | 0 | 0.08 |
| N4552 | 35 | 4 | 2 | 0 | 0.15 |
| N4697 | 18 | 2 | 0 | 0 | 0.10 |
| | 206 | 21 | 6 | 1 | 0.12 |

a. $L_X$ in unit of $10^{38}$ erg s$^{-1}$
b. This bin ($L_X > 20$) indicates the number of potential ULXs.

The luminous LMXB fractions ($F_{LL}$) are 0.22 (54 out of 242) and 0.12 (27 out of 233) in the young and old elliptical galaxy samples, respectively. While the total number of LMXBs ($L_X > 10^{38}$ erg s$^{-1}$) is similar in the two samples, the number of luminous LMXBs ($L_X > 5 \times 10^{38}$ erg s$^{-1}$) is higher by a factor of 2 in the young sample compared to the old sample. The proportional test (available in the *R* package, http://www.r-project.org/) results in a p-value of 0.001417 which corresponds to a confidence interval of 0.998, indicating that the difference is statistically significant at a 3.1σ confidence.

Figure 2 shows the $F_{LL}$ for each galaxy plotted against age; the collective $F_{LL}$ for each samples are also shown. To properly treat a small number of luminous sources from individual galaxies, we apply the Bayesian estimation technique developed for X-ray hardness ratios by Park et al. (2006) and estimate $F_{LL}$ and its error at 68% significance. The fraction ($F_{LL}$) of an individual galaxy is always consistent with the sample mean within the statistical error, except one young elliptical galaxy (NGC 3585), whose $F_{LL}$ is comparable with that of old ellipticals. $F_{LL}$ is clearly higher in young elliptical galaxies than in old elliptical galaxies. We also investigated possible dependencies of $F_{LL}$ on the stellar luminosity of the galaxy and on the GC specific frequency, but we found no clear trend. (see more discussions in Section 5).

It is also interesting to note that the young E sample hosts more (~5 times) ULX type LMXBs with $L_X > 2 \times 10^{39}$ erg s$^{-1}$ (see Table 4). However, the number of sources is too low to determine the statistical significance.

# 4. Discussion

We have investigated the effect of stellar age on the LMXB properties of elliptical galaxies by selecting seven young and seven old galaxies which were observed with Chandra deeply enough to detect LMXBs with Lx $\geq 10^{38}$ erg s$^{-1}$ (at a 90% detection limit). The average stellar ages of elliptical galaxies have been measured with optical spectroscopic, photometric and imaging observations providing important information to understand the evolution of these galaxies, such as rejuvenation by mergers (e.g., Trager et al. 2000; Thomas et al. 2005). Since the age measurements may be subject to significant uncertainties for some galaxies (e.g., Idiart et al. 2007; Rogers et al. 2009; Trager & Somerville 2009), we have considered only the average properties of the two samples. We take the young elliptical galaxy sample as a group where on average the secondary star formation episode occurred rather recently (later than roughly the half of the Hubble time), as opposed to a typical old, passively evolving stellar systems where the initial star formation ended long ago.

Chandra observations of various types of galaxies have shown that the XLF changes from old stellar systems to young stellar systems with the XLF slope getting flatter from elliptical (Kim & Fabbiano 2004; Gilfanov 2004) to spiral (e.g., Kong et al., 2003; Kilgard et al. 2005) and star burst galaxies (e.g., Zezas and Fabbiano 2002): $\beta \sim 2.0$ in elliptical galaxies (when fitted by a single power-law in a differential form, of $dN/dL_X$ vs. $L_X^{-\beta}$), while $\beta \sim 1.5$ in spiral and star burst galaxies.

A more careful look at elliptical galaxies (KF04; Gilfanov 2004; Voss & Gilfanov 2006; Kim et al. 2009) shows that the LMXB XLF is more complicated than a single power-law, having a high luminosity break at $L_X = 5 \pm 1.6 \times 10^{38}$ erg sec$^{-1}$ (KF04; Gilfanov 2004) and a low-luminosity break at $L_X$ = a few x $10^{37}$ erg sec$^{-1}$ (Voss & Gilfanov 2006; Kim et al. 2009; Voss et al. 2009). The situation at low luminosity is somewhat complicated, since there are reported differences between the XLFs of LMXBs in globular cluster (GC) and in the stellar field. In particular, there is a general lack of faint LMXBs in GCs when compared to field LMXBs (Kim et al. 2009; Voss et al. 2009). However, the XLF at the high luminosity is the same for field and GC LMXBs, at least for typical old elliptical galaxies (Kim E. et al 2006). The high luminosity break at $L_X = 5 \times 10^{38}$ erg sec$^{-1}$ is likely due to the different contribution from neutron star (NS) and black hole (BH) binaries, as suggested by Sarazin et al. (2001). Above the break ($L_X > 5 \times 10^{38}$ erg sec-1), the XLF slope becomes very steep ($\beta \sim 3$) and the very luminous X-ray sources (or ULX with $L_X > 2 \times 10^{39}$ erg sec$^{-1}$) may not even exist in typical old ellipticals, where the number of these sources is consistent with chance detection of background AGNs (Irwin et al. 2004). Instead, the typical XLF in star forming galaxies is flatter ($\beta \sim 1.6$) than that of elliptical galaxies. This flat XLF continues to higher luminosities (to the ULX $L_X$ range) and the XLF normalization appears to be linked with the galaxy's star formation rate (see review by Fabbiano 2006).

The main goal of this paper is to search for a possible age effect on the LMXB properties. So far there have been only two young elliptical galaxies observed with Chandra with exposures deep enough to detect a large number of LMXBs (NGC 720 and NGC 5018), that have been studied. These observations have led to the discovery of flat XLFs. In NGC 720, Jeltema et al. (2003) measured a XLF slope $\beta_{low} \sim 1.4 \pm 0.3$ for $L_X < 2 \times 10^{39}$ erg sec$^{-1}$ (i.e., without ULXs), which is considerably flatter than that of the typical old elliptical galaxy XLF. KF04 also find that NGC 720 (the youngest among their 14 early type galaxy sample) has the flattest slope ($\beta_{low}$ ~ 1.6, see Figure 2 in KF04). NGC 720 also hosts a large number (eight after correcting for a new distance) of ULX candidates with $L_X > 10^{39}$ erg sec$^{-1}$. Although two of them are likely background AGNs (Lopez-Corredorira & Gutierrez 2006), it is statistically unlikely that all of them are AGNs. In the young elliptical galaxy NGC 5018 (1.5 Gyr at 30 Mpc), Ghosh et al. (2005) reported six non-nuclear luminous LMXBs (with $L_X > 10^{39}$ erg sec$^{-1}$), but the total number of detected LMXBs (11) is too small to constrain the XLF.

Using carefully selected samples of young and old elliptical galaxies, we show that young elliptical galaxies are indeed intermediate in their XLF property between old elliptical galaxies and star forming galaxies. The XLF in young ellipticals continues without a significant break (at $L_X = 5 \times 10^{38}$ erg sec$^{-1}$), or possibly breaks at higher luminosity (at $L_X \sim 10^{39}$ erg s$^{-1}$). Our finding may suggest that the population of X-ray luminous BH binaries (including ULXs) is a strong function of age (e.g., Belczynski et al. 2004). If LMXBs are mainly formed in globular clusters (GCs), a large number of young GCs formed after the last merger (e.g., Schweizer 1987; Ashman & Zepf 1992; also review by Brodie and Strader 2006) might have triggered a number of luminous LMXBs. As the binary fraction in GCs is a strong function of time and decreases rapidly in the first 5 Gyrs (Ivanova et al. 2005), a higher X-ray binary fraction is expected in younger ellipticals for a given optical luminosity (or GC specific frequency) than old ellipticals. This effect will be enhanced because the second generation GCs are expected to be primarily red metal-rich (e.g., Schweizer 2003) and because red GCs are known to host the majority of GC-LMXBs (Kundu et al. 2002; Sarazin et al. 2003; Kim, E. et al. 2006).

We investigated if our result may be due to factors other than age. We can exclude dependencies on morphological type (E vs. S0), galaxy size, GC specific frequency ($S_N$) and kinematical property (fast vs. slow rotators). Although young galaxies tend to be S0's rather than E's, only one true S0 (NGC 4382) is included in our young galaxy sample. While the ranges of optical luminosities overlap (see Figure 3a), the young sample (blue circles) contains a few more luminous galaxies than the old sample (red squares). However, there is no trend between the fraction of luminous LMXBs ($F_{LL}$) and $L_K$ in the entire sample, or separately in the young and old sample. The only clear trend is that among those with similar $L_K$ (> $2 \times 10^{11}$ L$_\odot$), young galaxies have higher $F_{LL}$ than old galaxies. Since GC is one of the key factors in LMXB properties (e.g., Kim et al. 2009; Voss et al. 2009), we tested whether our result is affected by $S_N$ (in Figure 3b). Apparently there seems to be a weak trend between $F_{LL}$ and $S_N$. However, that is likely driven by age (as a primary parameter), because most old galaxies in our sample have higher $S_N$. In either the young (blue circles) and old (red squares) sample, we find no trend between the fraction of luminous LMXBs ($F_{LL}$) and $S_N$. Applying partial correlation tests (Pearson, Spearman and Kendall) to 11 galaxies with available $S_N$, we find considerably higher p-values (0.2-0.36) between $F_{LL}$ and $S_N$ for given age than those (0.01-0.09) between $F_{LL}$ and age for given $S_N$. Again, the only clear trend is that young galaxies tend to have higher $F_{LL}$ than old galaxies, regardless of $S_N$. Finally, young galaxies are often associated with fast rotators (as studied in the SAURON project, e.g., Cappellari, M., et al. 2007; Emsellem et al. 2007). Among our galaxies, three young (N720, N3377, N4382) and five old galaxies (N3379, N4278, N4374, N4472, N4552) are in the SAURON survey. While all three young galaxies are fast rotators, the old galaxies include both fast and slow rotators (N3379 and N4278 being fast rotators).

In summary, we found that the XLFs of LMXBs are different (with a statistical significance of 3σ) between two samples of young and old elliptical galaxies such that the young galaxy sample hosts more luminous LMXBs. The XLF of the young galaxy sample is intermediate between that of typical old elliptical galaxies and that of star forming galaxies. We speculate that the excess number of luminous LMXBs in 'young' elliptical galaxies may be the result of binary formation stimulated by the recent major/minor mergers responsible for the rejuvenation of the stellar population of these galaxies. It is important to know whether those luminous LMXBs found in young E's are primarily associated with globular clusters or they are in the field. However, we cannot address this question, because of limited GC data in the young sample. While we are confident of our observational results, a detailed theoretical explanation for this X-ray rejuvenation effect is beyond the scope of this work. We hope that our results may stimulate new work in this regard.

To illustrate the current status of the XLF of LMXBs, we show in Figure 4 a schematic diagram. In the intermediate luminosity range ($L_X = 5 \times 10^{37} - 5 \times 10^{38}$ erg s$^{-1}$), where most LMXBs in nearby galaxies (10-20 Mpc) are detected in typical Chandra observations (for 30-40 ksec), the XLF is well represented by a single power-law ($dN/dL_X \sim L_X^{-2}$ or $dN/dlnL_X \sim L_X^{-1}$) (KF04; Gilfanov 2004), regardless of their association with GCs and their host galaxy ages. At the lower luminosities ($L_X < 5 \times 10^{37}$ erg s$^{-1}$), where LMXBs are detected in only a small number of galaxies with deep Chandra observations, the XLF starts to differ between GC and field LMXBs, GC-LMXBs having significantly fewer faint LMXBs (Kim et al. 2009; Voss et al. 2009). At $L_X \sim 5 \times 10^{37}$ erg s$^{-1}$, the XLF may have a bump which is more pronounced in the field LMXBs than in the GC LMXBs, likely due to the NS binaries with red giant donors (Kim et al. 2009). At high luminosities ($L_X > 5 \times 10^{38}$ erg s$^{-1}$), only a small number of LMXBs are detected in individual galaxies, particularly in typical old elliptical galaxies (e.g., see Table 4). In this Lx range, the young galaxy sample hosts more luminous LMXBs, XLF being intermediate between that of typical old elliptical galaxies and that of star forming galaxies (this paper).

**ACKNOWLEDGEMENTS**

The data analysis was supported by the CXC CIAO software and CALDB. We have used the NASA NED and ADS facilities, and have extracted archival data from the *Chandra* archives. This work was supported by the *Chandra* GO grant G08-9133X (PI: Kim). We thanks Tassos Fragos and Vicky Kalogera for interesting conversations.

**Figure Captions**

Fig 1. Comparison of X-ray luminosity functions of LMXBs: young (top) and old (bottom) elliptical galaxy samples. The XLFs are fitted with a single power-law (left) and a broken power-law (right). The best fit models before (green histogram) and after (red histogram) completeness correction are also plotted. Two diagonal lines with a slope of 1 (or $\beta=2$ in the differential XLF form) are drawn for visibility. The red vertical lines indicate the break luminosity in the broken power-law fit. The three green vertical lines indicate $L_X = 5 \times 10^{38}$, $10^{39}$ and $2 \times 10^{39}$ erg s$^{-1}$ (from right to left) to make the different break luminosities of the young and old samples more visible.

Fig 2. The fraction ($F_{LL}$) of luminous ($L_X > 5 \times 10^{38}$ erg s$^{-1}$) LMXBs against age. The big blue circle and red square indicate mean values of $F_{LL}$ for young and old elliptical galaxies, respectively.

Fig 3. The fraction ($F_{LL}$) of luminous ($L_X > 5 \times 10^{38}$ erg s$^{-1}$) LMXBs against (a) $L_K$ and (b) $S_N$. The blue circles and red squares indicate young and old elliptical galaxies, respectively.

Fig 4. A schematic view of the X-ray luminosity function of LMXBs.

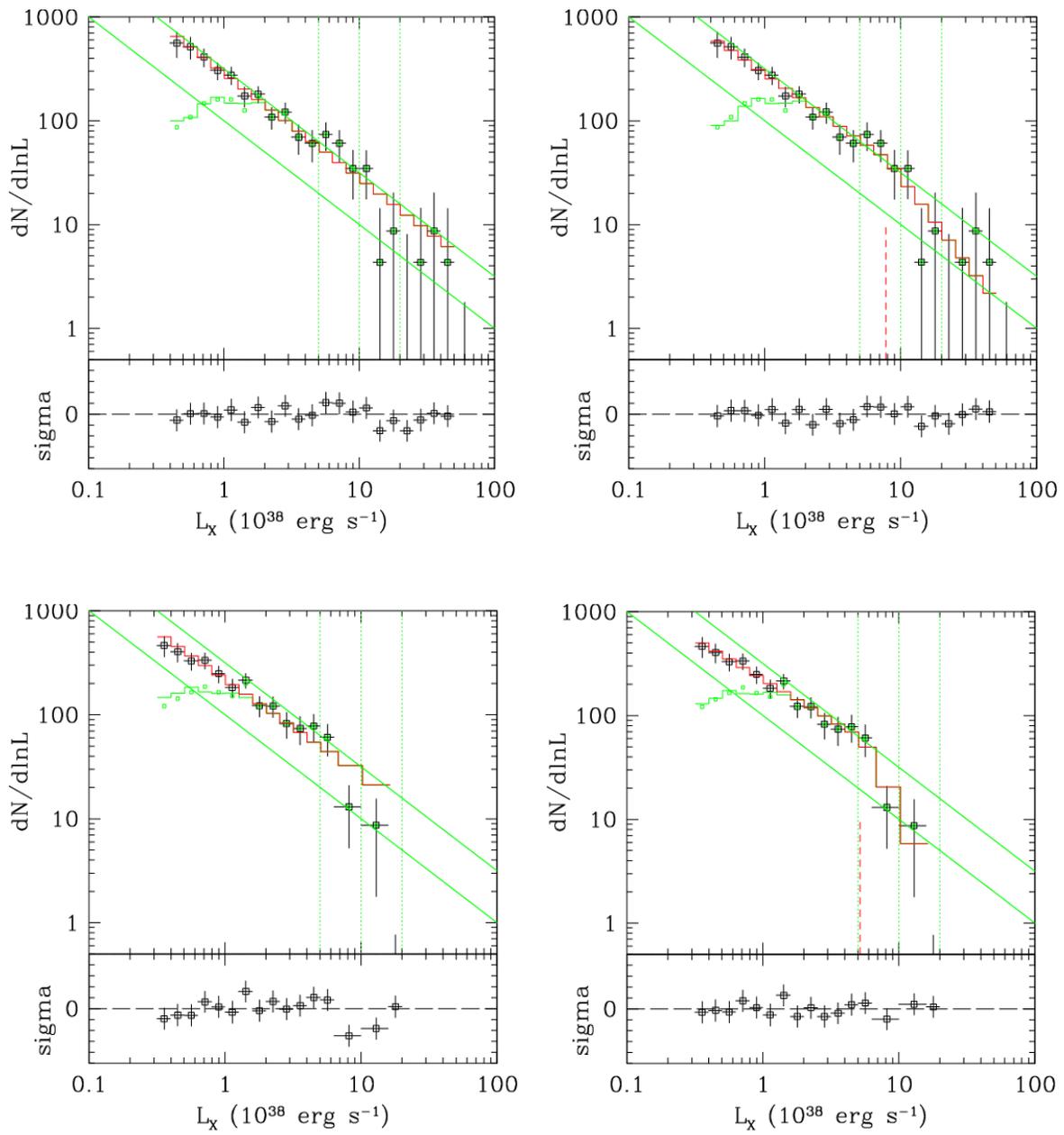

Fig 1.

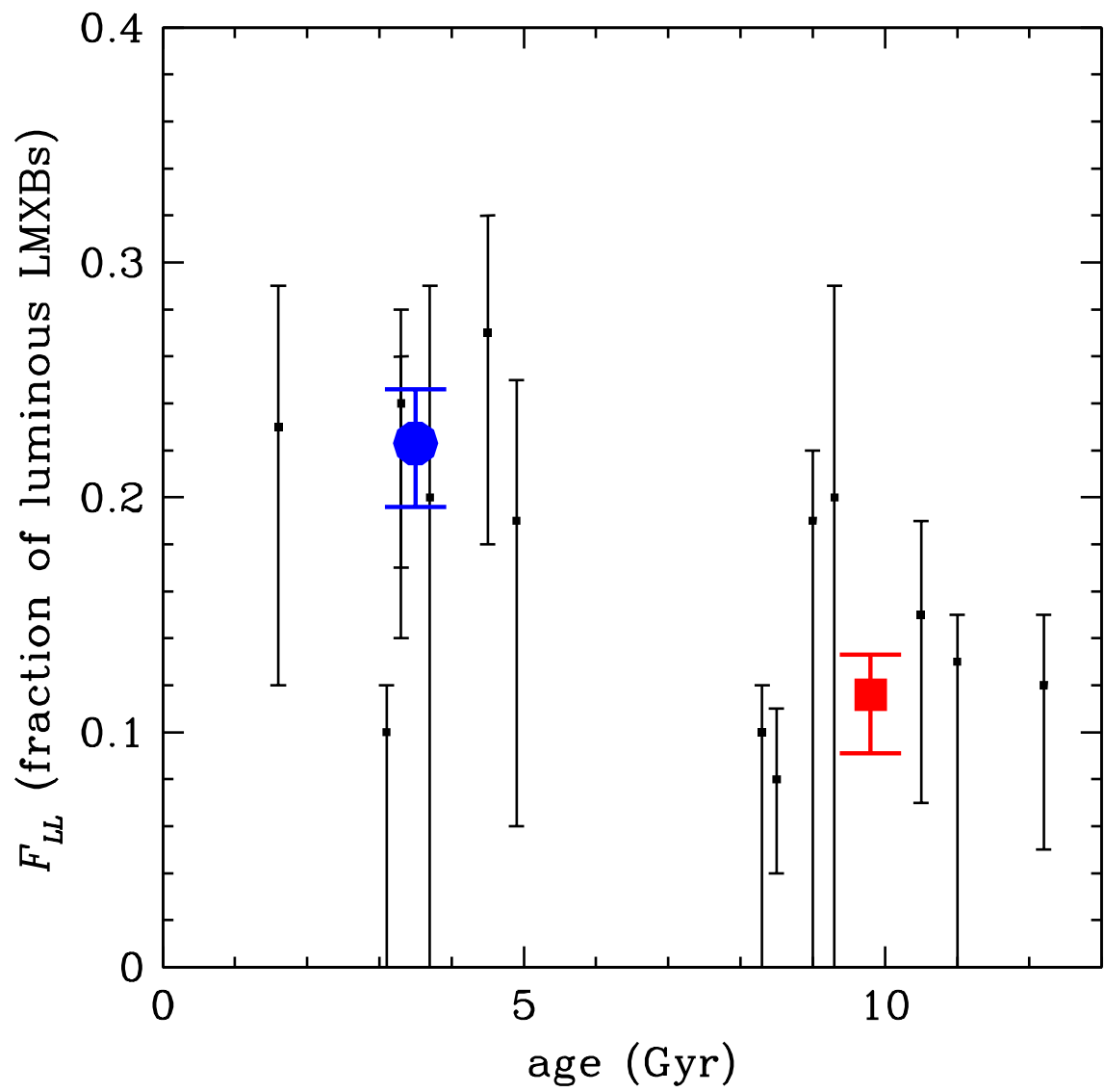

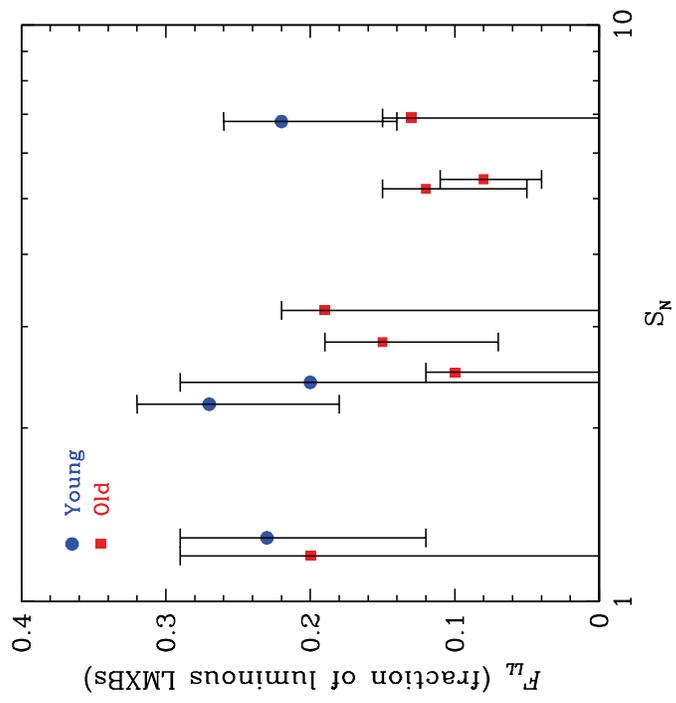
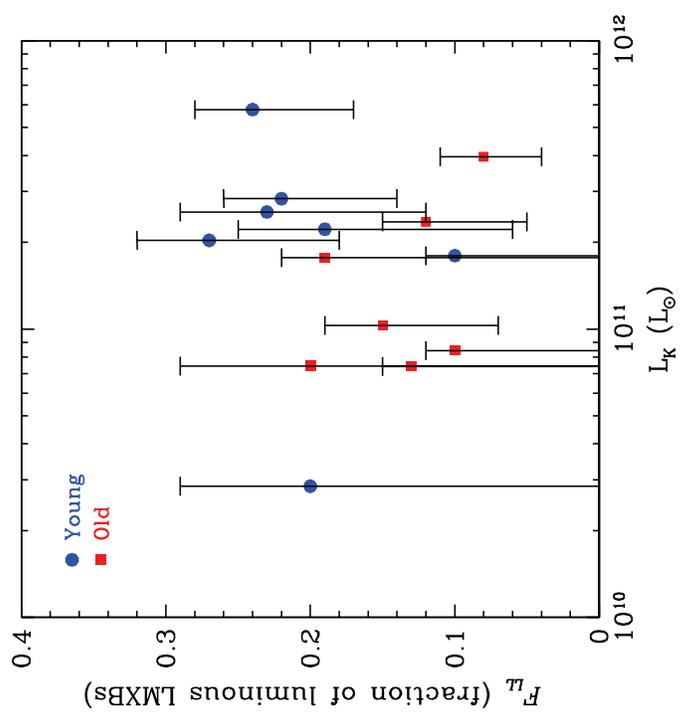

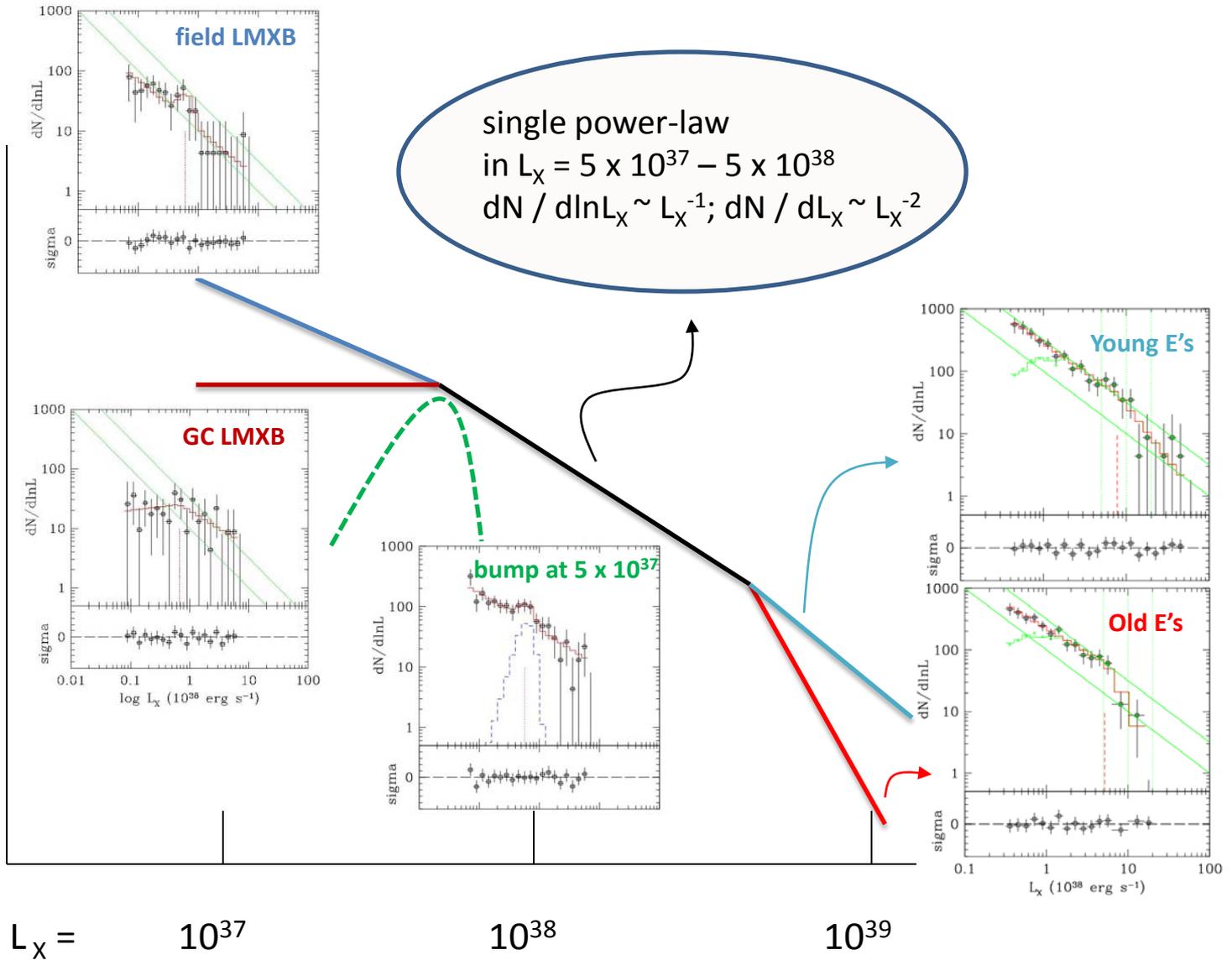